\documentclass[12pt,preprint]{aastex}

\slugcomment{Accepted for publication in The Astronomical Journal}

\begin{document}
\title{{\it HST}/ACS Images of the GG Tauri Circumbinary Disk}
\author{J.E. Krist\altaffilmark{1,2},
K.R. Stapelfeldt\altaffilmark{1},
D.A. Golimowski\altaffilmark{3},
D.R. Ardila\altaffilmark{3},
M. Clampin\altaffilmark{4},
A.R. Martel\altaffilmark{3},
H.C. Ford\altaffilmark{3},
G.D. Illingworth\altaffilmark{5},
G.F. Hartig\altaffilmark{2}}

\altaffiltext{1}{Jet Propulsion Laboratory, 4800 Oak Grove Dr., Pasadena, CA 91109.}
\altaffiltext{2}{Space Telescope Science Institute, 3700 San Martin Dr., Baltimore, MD 21218.}
\altaffiltext{3}{Department of Physics and Astronomy, Johns Hopkins University, 3400 North Charles Street, Baltimore, MD 21218.}
\altaffiltext{4}{NASA Goddard Space Flight Center, Code 681, Greenbelt, MD 20771.}
\altaffiltext{5}{UCO/Lick Observatory, University of California, Santa
Cruz, CA 95064.}

\begin{abstract}

{\it Hubble Space Telescope} Advanced Camera for Surveys images of the young
binary GG Tauri and its circumbinary disk in $V$ and $I$ bandpasses were
obtained in 2002 and are the most detailed of this system to date .  They
confirm features previously seen in the disk including: a ``gap'' apparently
caused by shadowing from circumstellar material; an asymmetrical distribution
of light about the line of sight on the near edge of the disk; enhanced
brightness along the near edge of the disk due to forward scattering; and a
compact reflection nebula near the secondary star.  New features are seen in
the ACS images: two short filaments along the disk; localized but strong
variations in the disk intensity (``gaplets''); and a ``spur'' or filament
extending from the reflection nebulosity near the secondary.  The back side of
the disk is detected in the $V$ band for the first time.  The disk appears
redder than the combined light from the stars, which may be explained by a
varied distribution of grain sizes.  The brightness asymmetries along the disk
suggest that it is asymmetrically illuminated by the stars due to extinction by
nonuniform circumstellar material or the illuminated surface of the disk is
warped by tidal effects (or perhaps both).  Localized, time-dependent
brightness variations in the  disk are also seen. 

\end{abstract}


\section{Introduction}

Judging by population statistics in nearby star forming regions like the
Taurus, Ophiuchus, and Chameleon molecular clouds (Duch{\^ e}ne 1999), the
majority of young ($<$10 Myr) stars in them belong to bound, multiple systems.
Many, if not most, of these stars are surrounded by dusty envelopes or disks.
Some circum{\it stellar} disks (disks around individual stars) in multiple
systems have been studied from visible through radio wavelengths ({\it e.g.} HK
Tau/c; Stapelfeldt et al. 1998).  Because of tidal interactions with the other
stars in their systems, these stars may have disks truncated at outer radii.
Circum{\it binary} disks, however, appear to be somewhat more rare, and only
two, UY Aurigae (Close et al. 1998) and GG Tauri, have been studied thoroughly
from optical to radio wavelengths.  

GG Tau is a young, multiple system located in the Taurus star forming region,
140 pc distant.  It includes two binaries, the brighter of which, GG Tau Aa/Ab
(hereafter simply called GG Tau), is composed of two T Tauri stars separated by
0\farcs 25.  The brighter star, GG Tau Aa, was given an M0 spectral
classification by Hartigan \& Kenyon (2003) and the fainter GG Tau Ab is type
M2.  This system is surrounded by a fairly massive (0.1 M$_{\odot}$),
optically-thick disk.  Imaging at radio (Guilloteau, Dutrey, \& Simon 1999),
near-infrared (Silber et al. 2000; Roddier et al.  1996), and visible (Krist,
Stapelfeldt, \& Watson 2002) wavelengths has revealed that inside of 150 AU the
disk has been largely cleared by tidal interactions with the binary.  The dust
near the inner edge is illuminated by the stars, creating the appearance of a
narrow circumbinary ring inclined by $\sim37^{\circ}$ from face-on.  Dust at 
larger radii may be shadowed by the inner material, so the disk is likely
larger than it appears in scattered light.

The first adaptive optics (AO) images of GG Tau by Roddier et al. (1996;
hereafter R96) showed the ring in reflected light in {\it JHK} bands.  It
appeared clumpy and had radial spokes that seemed to connect the ring to the
stars.  R96 suggested that these could be accretion streams, similar to those
predicted by Artymowicz \& Lubow (1996).  These features, however, appeared
nearly as bright as the optically thick disk, so they would have to be rather
dense to reflect a sufficient amount of light to be detected.
 
{\it Hubble Space Telescope} ({\it HST}) imaging of GG Tau at visible
wavelengths (Krist, Stapelfeldt, \& Watson 2002, hereafter KSW02) using the
WFPC2 camera and in the near-IR with NICMOS (Silber et al.  2000; McCabe,
Duch{\^ e}ne, \& Ghez 2002, hereafter MDG02) showed that the ring is generally
smooth and lacks radial spokes, suggesting that those seen by R96 were
instrumental artifacts.  The {\it HST} images revealed brightness and
structural asymmetries in the ring.  For instance, there appears to be a
``gap'' in the ring that is likely a shadow caused by material between the ring
and the stars.  Also, the brightest section of the ring is not along the
closest edge, as would be expected from forward scattering by dust grains in a
uniformly illuminated, symmetrical disk, but is instead offset in azimuth by
$\sim20^{ \circ}$.  These and other asymmetries suggest that the illuminated
surface of the disk is warped and/or the illumination is shadowed by
circumstellar material.  The latter case is bolstered by a compact reflection
nebulosity seen near the secondary star in the WFPC2 image.  The {\it HST}
studies also showed that the ring is redder than the combined observed light
from the stars.

Because the ring is optically thick, the dust density distribution and grain
properties cannot be directly derived from the reflected-light images.
Three-dimensional, multiple-scattering models have been used to examine
possible dust distributions and grain properties (Wood, Crosas, \& Ghez 1999;
Duch{\^e}ne et al. 2004, hereafter D04) with constraints provided by both the
scattered light and thermal emission images.  The two important measurements
provided by the visible and near-IR images are the ratio of the forward side to
back side (F/B) brightnesses and the ratio of integrated ring flux to the
combined stellar fluxes.  The first ratio sets limits on the degree of forward
scattering in the models, which is a function of particle size.  The variation
in forward scattering with wavelength can therefore place constraints on the
assumed grain size distribution function.  The second ratio can indicate the
amount of light from the stars lost by multiple scattering or extinction by
circumstellar material. 

Until now, the only images of the GG Tau ring at visible wavelengths were those
from the WFPC2 camera on {\it HST}.  However, the F555W (WFPC2 $V$ band) image
was neither deep enough nor sufficiently free of point spread function (PSF)
subtraction artifacts to unambiguously show the back side of the ring, and thus
failed to provide a strong constraint on the variation of forward scattering
with wavelength.  To remedy this situation, images of GG Tau have been taken
using the Advanced Camera for Surveys (ACS) on {\it HST}.  Its finer pixel
scale and more stable PSF allows better subtractions of the stellar glare,
which are required to reveal the ring against the background glare from the
stars.

\section{ Observations, Stellar Photometry, and PSF Subtraction} 

The GG Tau images were taken as part of the ACS Investigation Definition Team's
circumstellar disk program ({\it HST} program 9295) on 12 September 2002.  The
system was observed with the ACS High Resolution Camera (HRC), which has a
pixel scale of $\sim$0\farcs 025 pixel$^{-1}$ and a PSF FWHM of 50 mas in the
$V$ band.  Identical sets of exposures were taken at two orientations with the
telescope rolled about its optical axis by 28$^{\circ}$.  These rolls permit
improved differentiation of PSF artifacts (which remain static as seen by the
detector regardless of roll) from real disk structures (which change
orientation with telescope roll).  They also allow regions of the ring obscured
by the diffraction spikes at one roll to be replaced by unobscured pixels from
the other.  At each roll, identical sets of short-to-long duration exposures
were taken through filters F555W (ACS $V$ band) and F814W (ACS $I$ band) (Table
1).  The data were calibrated by the {\it HST} data archive.  Duplicate
exposures were combined with cosmic ray rejection and saturated pixels in long
exposures were replaced by scaled values from shorter ones, providing
high-dynamic-range images.

Because of the faintness of the GG Tau disk and its proximity to the stars, the
stellar PSFs had to be subtracted using images of another star.  HD 260655
(K7V) was chosen as a reference PSF because its color was similar to that of
the brighter binary component, GG Tau Aa.  It was observed in an orbit
immediately following the GG Tau observations in order to reduce the influence
of time-dependent focus changes that could lead to PSF mismatches and increased
subtraction residuals.  Short-to-long exposures were taken and processed in the
same manner as the GG Tau images.  Unfortunately, the short F814W image was
slightly saturated.

\begin{deluxetable}{lcl}
\footnotesize
\tablecaption{ACS GG Tau and Reference PSF Exposures\label{tbl1}}
\tablewidth{0pt}
\tablehead{ \colhead{Star} & \colhead{Filter} & \colhead{Exposure} }
\startdata
   GG Tau$^1$ &  F555W      &      2 x 3s \\
              &             &      2 x 30s \\
              &             &      2 x 1275s \\
              &  F814W      &      2 x 1s \\
              &             &      1 x 10s \\
              &             &      2 x 25s \\
              &             &      2 x 250s \\
              &             &      \\
   HD 260655  &  F555W      &      2 x 0.2s \\
              &             &      2 x 0.8s \\
              &             &      2 x 2s \\
              &             &      2 x 10s \\
              &             &      2 x 50s \\
              &             &      3 x 300s \\
              &  F814W      &      2 x 2s \\
              &             &      2 x 50s \\
              &             &      2 x 200s \\
              &             &      3 x 830s
\enddata
\tablenotetext{1}{The listed GG Tau exposures were duplicated after
rolling the telescope 28$^{\circ}$.}
\end{deluxetable}

\subsection{PSF Fitting and Stellar Photometry}

Because the separation between the binary components is small and portions of
their PSFs overlap, simple aperture photometry may introduce unacceptable
errors.  To obtain accurate positions and flux measurements, simultaneous
PSF fitting of both components was undertaken.  For the F555W data, Tiny Tim
model PSFs (Krist \& Hook 2003) were matched to the observations, as they were
in the WFPC2 study (KSW02).  Iterative model-fitting phase retrieval analysis
(Krist \& Burrows 1995) was used to determine the aberration parameters (focus,
coma, astigmatism) appropriate for the observations by measuring GG Tau Aa in
the shortest exposure at the first orientation.  These values were then used in
Tiny Tim to generate subsampled, geometrically-distorted models with colors
similar to the GG Tau stars.  Within an interative optimization routine the
models were shifted via cubic convolution interpolation, rebinned to normal
pixel sampling, convolved with a kernel representing blurring from HRC CCD
charge diffusion, and then scaled in intensity to match the binary components.
The {\it HST} images were not corrected at this point for geometric distortion
to avoid potential errors caused by interpolation.  

PSF fitting photometry in F814W was more complicated because the HRC CCD
suffers from a ``red halo'' (Sirriani et al. 1998).  This appears as a
diffuse, smooth halo of light that can be seen extending out to $\sim$4'' from
the stars in the deep F814W exposures.  Its surface brightness decreases with
increasing radius.  It is caused by red photons that pass completely through
the CCD and are scattered to large angles within the glass mounting substrate
and then back into the detector.  The halo begins to appear between
$6000-7000$\AA\ and by 10000\AA\ it can contain more than 30\% of the stellar
flux.  It is seen in the unsubtracted F814W image in Figure 1, where it washes
out the diffraction structure in the wings.  The effect is sensitive to the
color of the object over a broad passband, leading to potential mismatches
between target and reference PSFs, even for similar spectral types.

\begin{figure}
\plotone{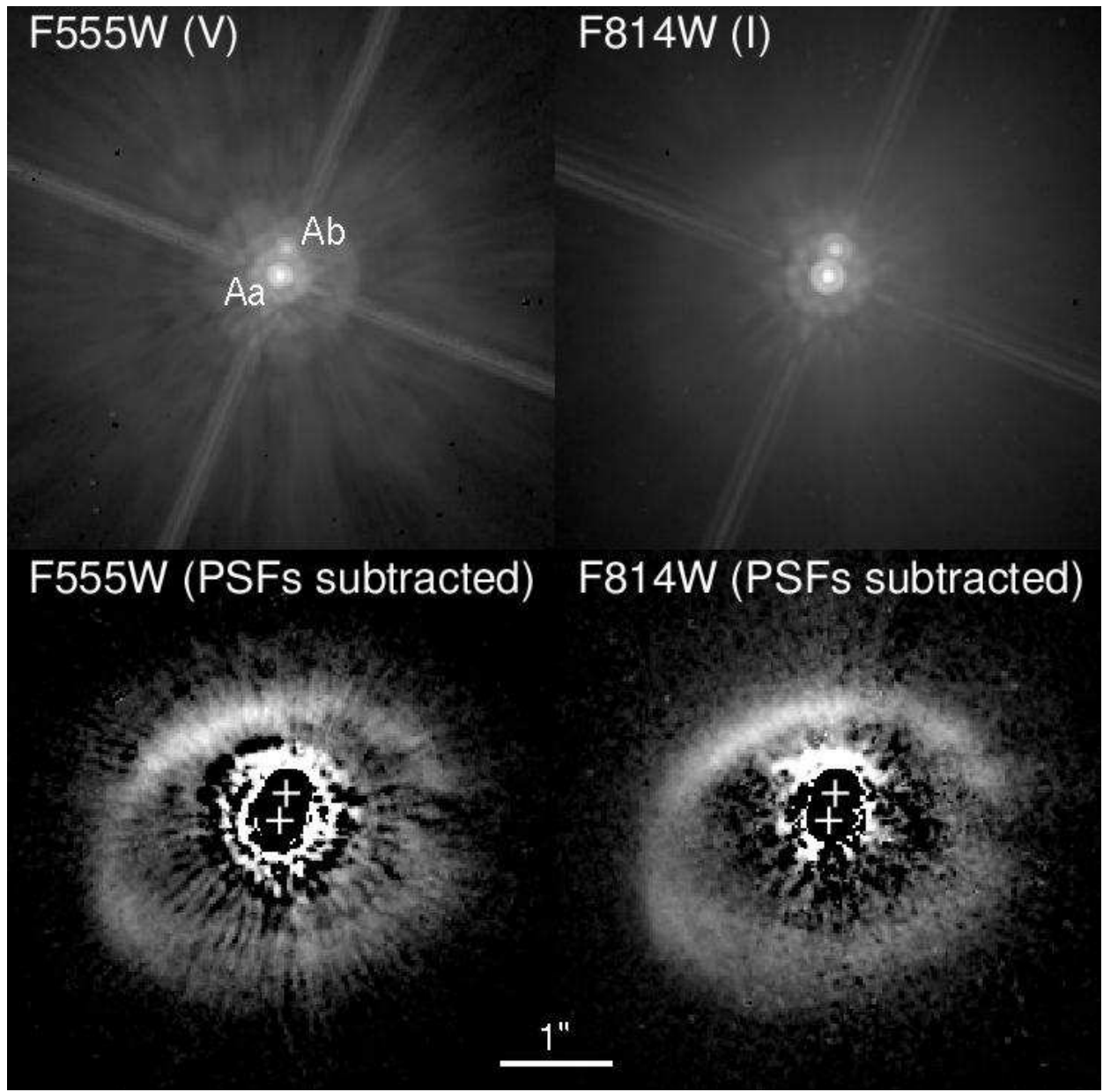}
\caption{ACS images of GG Tau. (Top) Unsubtracted images in filters F555W ($V$)
and F814W ($I$), displayed with logarithmic stretches.  (Bottom) Combined first
and second orientation images after subtraction of reference PSFs, displayed
with square-root stretches.  The locations of the binary components are marked
with crosses.  North is up in all of these images.} 
\label{fig1} 
\end{figure}

At this time, Tiny Tim does not model the red halo with sufficient accuracy to
provide good subtractions.  Rather than fitting the F814W GG Tau binary with
PSF models we instead used an image of HD 17637 that was observed in an {\it
HST} calibration program (we could not use HD 260655 as it was saturated in
this filter).  HD 17637 is a K5V star, so it is slightly more blue than GG Tau
Aa.  However, it should be a better match than a Tiny Tim model.  The PSF
image, which was not corrected for geometric distortion, was interpolated to
$5\times$ finer sampling (the observed HRC PSF is oversampled by a factor of
1.4 at $\lambda=8000$\AA\ relative to Nyquist sampling).  It was then used by
the same PSF fitting routine described previously (excluding charge diffusion
blurring, which the observed reference PSF already contained).  The fluxes were
aperture corrected.  Because the stellar photometry was obtained in the
distorted images, adjustments to the fluxes were required to account for the
difference between the pixel area assumed in the calibrated data and the true
area (this is described in the ACS Data Handbook; Pavlovsky et al. 2004).  The
average fluxes of the measurements at each roll were converted to standard $V$
and $I_c$ magnitudes (Table 2) using the SYNPHOT synthetic photometry package.
An M1.5V spectrum was assumed for both components of GG Tau A.  The estimated
photometric errors are $\pm$0.03 mag.

The positions of the binary components were corrected for geometric distortion,
and the resulting orientations were converted to position angles based on the
telescope orientations provided in the image file headers.  The mean
separations are 0\farcs 251 (F555W) and 0\farcs 250 (F814W), with errors of
$\pm$0\farcs 002.  The mean position angles are 346.3$^{\circ}$ (F555W) and
345.9$^{\circ}$ (F814W), with errors of $\pm0.5^{\circ}$.  

The flux of the ACS reference star, HD 260655, was also measured in each filter
to derive the normalization factors for PSF subtractions.  The F555W flux was
obtained by fitting the Tiny Tim PSF.  As with GG Tau, the fit in F814W was
done using HD 17637.  HD 260655 was saturated in the core in F814W, and those
pixels were given zero weight in the fit.  As with GG Tau, the fluxes were
converted to standard magnitudes assuming a M1.5V spectrum.  The estimated
errors are $\pm0.03$ mag (F555W) and $\pm0.06$ mag (F814W). 

Besides GG Tau A, other stars are seen in the HRC field and were measured using
$r=0\farcs 4$ apertures in the distortion-corrected images.  The stars'
photometry and positions relative to GG Tau A are given in Table 2.  The fluxes
were aperture corrected and converted to standard magnitudes assuming a M5.5V
spectrum for the GG Tau B components and a K5V spectrum for the others.  The
estimated photometric errors are $\pm0.05$ for GG Tau B and $\pm0.09$ for the
others.  The GG Tau B binary has a separation of $\sim$1\farcs 5.  The fainter
component, Bb, is very red ($V-I_c=4.55$) and was assigned a spectral type of
M7 by White et al.  (1999).  The brighter star, Ba, was given an M5 type.
Designations for the other stars in the field have not been made in previous
studies, so we assign them arbitrary names.  The colors of these stars imply
that they are of types earlier than M2.  One of them, which we designate GG Tau
1, was first noted by Silber et al. (2000), and further observations by Itoh et
al.  (2002) indicated that its motion relative to GG Tau A is consistent with
it being a background star.  

For comparison with the 1997 WFPC2 observations (KSW02) and the new ACS ones,
the 1994 WFPC2 F555W and F814W images of GG Tau Aa/Ab (Ghez et al. 1997) were
also measured by fitting Tiny Tim PSF models using the KSW02 routines.  These
measurements are independent of those reported by Ghez et al. (who fitted
observed PSFs rather than models), which were later converted to standard
magnitudes by White \& Ghez (2001).  Our values for these images for Aa \& Ab
are $V$=12.34 $\pm0.03$ \& 15.31 $\pm0.06$ and $I_c$=10.50 $\pm0.03$ \& 12.28
$\pm0.04$.  These are about 0.05 mag (Aa) and 0.1 mag (Ab) fainter than the
values reported by White \& Ghez (2001), due either to the differences in
measurement techniques or assumptions for spectral types used when converting
to the standard magnitude system.  For consistency, we use only our photometric
measurements and those of KSW02.   There were no differences between our
measurements of the position angle and separation and those reported by Ghez et
al. (1997).  Comparison of the mean of the F555W and F814W ACS position angles
to the 1994 angle provides an average binary angular motion of $-1.52^{\circ}$
yr$^{-1} \pm0.1^{\circ}$.  There is no significant change in separation between
the two epochs.

\begin{deluxetable}{lrrcrr}
\footnotesize
\tablecaption{GG Tauri System Stellar Photometry\label{tbl2}}
\tablewidth{0pt}
\tablehead{ \colhead{Star} & \colhead{$V$} & \colhead{$I_c$} & 
\colhead{$V-I_c$} & \colhead{Sep$^a$} & \colhead{P.A.$^b$} }
\startdata
GG Tau Aa  & 12.23 & 10.43 & 1.80 &            & \\
GG Tau Ab  & 15.18 & 12.37 & 2.81 &            & \\
GG Tau Ba  & 17.11 & 13.41 & 3.70 & 10\farcs 2 & 185.6$^{\circ}$ \\
GG Tau Bb  & 19.94 & 15.39 & 4.55 & 11\farcs 2 & 179.8$^{\circ}$ \\
GG Tau 1   & 21.55 & 19.28 & 2.27 &  6\farcs 2 & 241.1$^{\circ}$ \\
GG Tau 2   & 16.90 & 16.14 & 0.76 & 13\farcs 5 & 211.5$^{\circ}$ \\
GG Tau 3   & 25.4  & 22.6  & 2.8  & 11\farcs 6 & 7.9$^{\circ}$ \\
GG Tau 4   & 16.95 & 16.24 & 0.71 & 16\farcs 7 & 45.9$^{\circ}$\\
GG Tau 5   & 22.45 & 20.27 & 2.19 & 16\farcs 7 & 42.2$^{\circ}$ \\
HD 260655  &  9.53 &  7.86 & 1.67 &            &
\enddata
\tablenotetext{a}{Distance measured from GG Tau A binary midpoint}
\tablenotetext{b}{Position angle measured from GG Tau A binary midpoint}
\end{deluxetable}

\subsection{PSF Subtraction}

Using the measured photometry and positions, synthetic binaries for each filter
at each orientation were constructed by shifting and intensity-scaling the HD
260655 images.  These images were subtracted from GG Tau, and the resulting
images were then corrected for geometric distortion.  The data for each filter
from both rolls were combined by rotating the image from the second roll to
match the first, replacing pixels within the diffraction spikes at one roll
with those from the other, and then averaging the two.  The results are shown
in Figure 1.  Because of the larger mismatch in color between the GG Tau stars
and HD 260655 within the F555W passband, the subtraction residuals (radiating
streaks and rings around the stars) are greater in that filter than in F814W.

An initial review of the F814W images revealed significant contamination by
residuals caused by the incomplete subtraction of the red halo (Figures 2a,2b).
This remnant halo may be caused by the color mismatch between GG Tau Ab
($V-I_c=2.79$) and the reference PSF ($V-I_c=1.77$).  However, given that GG
Tau Aa is $\sim6\times$ brighter than Ab, it may more likely be the result of
small differences between the spectra of Aa and HD 260655 within the F814W
passband, despite their nearly equal $V-I_c$ colors. 

\begin{figure}
\plotone{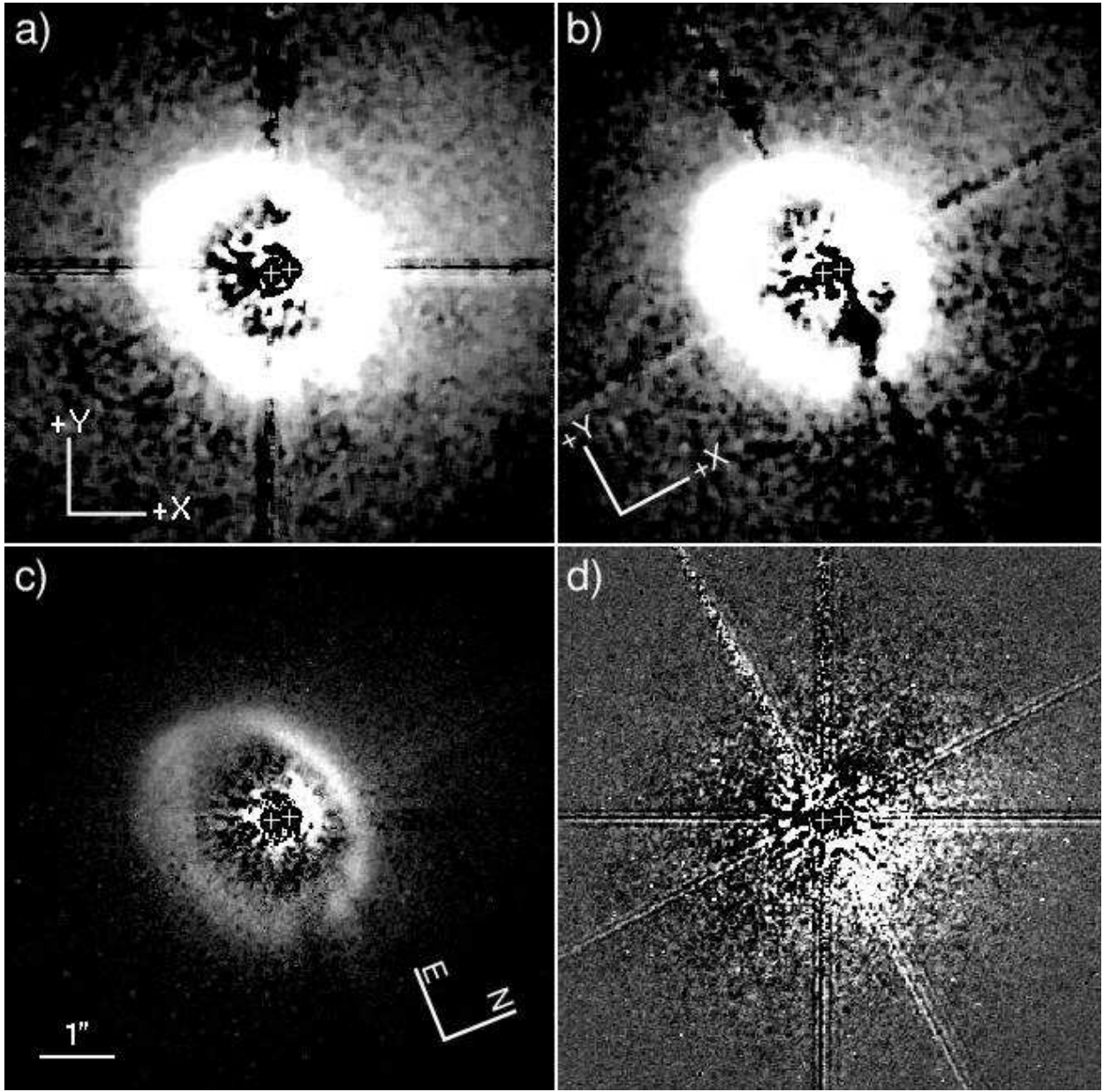}
\caption{ACS F814W ($I$) images of GG Tauri demonstrating the impact of the red
halo.  (a) PSF-subtracted image from the first orientation with an intensity
stretch chosen to highlight the residual halo, which can be seen outside of the
ring.  The halo is brighter towards the upper right. (b) The image from the
second orientation.  Note that the asymmetry in the halo is fixed with respect
to the detector axes (indicated by the +X,+Y compass). (c) Square-root stretch
of the combined F814W images, for reference. (d) The subtraction of the second
F814W PSF-subtracted image from the first, showing the instrumental halo
asymmetry (linear stretch).  All of the images are at the same orientation.}
\label{fig2}
\end{figure}
 
When the F814W PSF-subtracted, distortion-corrected images of GG Tau from each
roll are rotated to a common sky orientation and the second is subtracted from
the first, an asymmetry in the halo becomes apparent (Figure 2d).  It is
clearly instrumental because the residuals are not confined to the ring.  Some
of the asymmetry is caused by the distortion correction combined with the
different rolls.  The HRC detector is tilted relative to the incident beam to
minimize focus variations over the field, resulting in significant geometric
distortion (for instance, the diffraction spikes are 7$^{\circ}$ from being
perpendicular in the uncorrected images).  Assuming the halo is circularly
symmetric in detector coordinates, then after distortion correction it will
appear elliptical with the major axis roughly along the (+X,+Y) diagonal.  When
the second image is rotated to match the sky orientation of the first, the
angle of the residual halo axis will be different between the two images
(compare 2a and 2b).  If the halo was perfectly symmetric on the detector, then
the residuals should be symmetric as well, but they are not, suggesting some
intrinsic asymmetry.  Given that the detector is tilted relative to the
incoming beam, the halo may be brighter on one side than the other due to a
preferential scattering angle within the mounting substrate.  The halo may also
be field-dependent due to changes in the incident beam angle.

In regions where the halo effect is strong, we estimate that the portion of the
subtraction residuals caused by halo mismatches correspond to $\sim$15\% the
brightness of the disk.

\section{Results}

\subsection{General Features}

The PSF-subtracted ACS images provide the clearest views of the GG Tau disk
obtained so far in scattered light.  The use of two rolls provides complete
images of the ring and avoids diffraction spike residuals seen in the previous
single-roll WFPC2 and NICMOS images.  Excluding the red halo problem, the
relatively field-independent ACS PSF provides somewhat better subtractions than
are possible with WFPC2, resulting in higher contrast views of the disk
structure (Figure 3).  The finer resolution also permits better registration of
the PSFs with reduced interpolation artifacts.  In addition, the deeper
exposures allow the far, southern side of the ring to be unambiguously imaged
in $V$ for the first time.

In general, the system looks much as it did in the WFPC2 and NICMOS images,
though more subtle intensity variations are visible.  The outer apparent edge
measures 3\farcs 7 by 3\farcs 0.  The northern side is peak is $\sim3.3\times$
brighter than the southern.  This is expected from forward scattering by dust
grains, as the northern side is closer to us as determined from the measured
rotation of the disk and the motion of the binary (Guilloteau, Dutrey, \& Simon
1999).  It has maximum surface brightnesses of $V\approx18.4$ and
$I_c\approx15.9$ mag arcsec$^{-2}$.  The southern, more distant side of the
ring appears about twice as thick as the nearer one.  This is a projection
effect, as both the inner wall and upper surface of the backside of the disk
are seen there while in the north only the illuminated upper surface is visible
(see Figure 2 in Silber et al.  2000).  The west side appears thinner,
brighter, and more irregular than the east side.  The inner edge of the ring is
fairly well defined in the north but is diffuse to the east and south, and
along the west it is highly irregular.  The outer edge is fairly smooth except
in the west.  Because the disk is optically and geometrically thick, its
inclination causes the midpoint between the stars to appear offset
$\sim$0\farcs 25 forward from the ring center (as defined by the center of the
ellipse fit to a trace through the ring peak azimuthal isophotes).  

The images clearly show the 0\farcs 4 wide ``gap'' in the western section of
the ring between PA=$254^{\circ}-275^{\circ}$ that was identified in previous
studies (Figure 3).  Within it the surface brightness decreases by one-half
relative to the adjacent sections of the ring.  In $V$ a PSF subtraction
residual passes through the middle of the gap.  There are other, less prominent
``gaplets'' along the northwest section centered at PA$\approx290^{\circ}$ and
320$^{\circ}$.  There are two thin filaments along the outer edge of the ring
between PA=$100^{\circ}-150^{\circ}$ in a section previously called the
``kink'' (Silber et al. 2000).  They overlap with a separation of 0\farcs 11
and are seen in both $V$ and $I$ bands as well as in the WFPC2 $I$ band image.
They are not resolved in the NICMOS images.  The filaments are better defined
in $V$ than $I$, perhaps due to the higher resolution or an increase in the
opacity at shorter wavelengths in the dust lane separating them.

\begin{figure}
\plotone{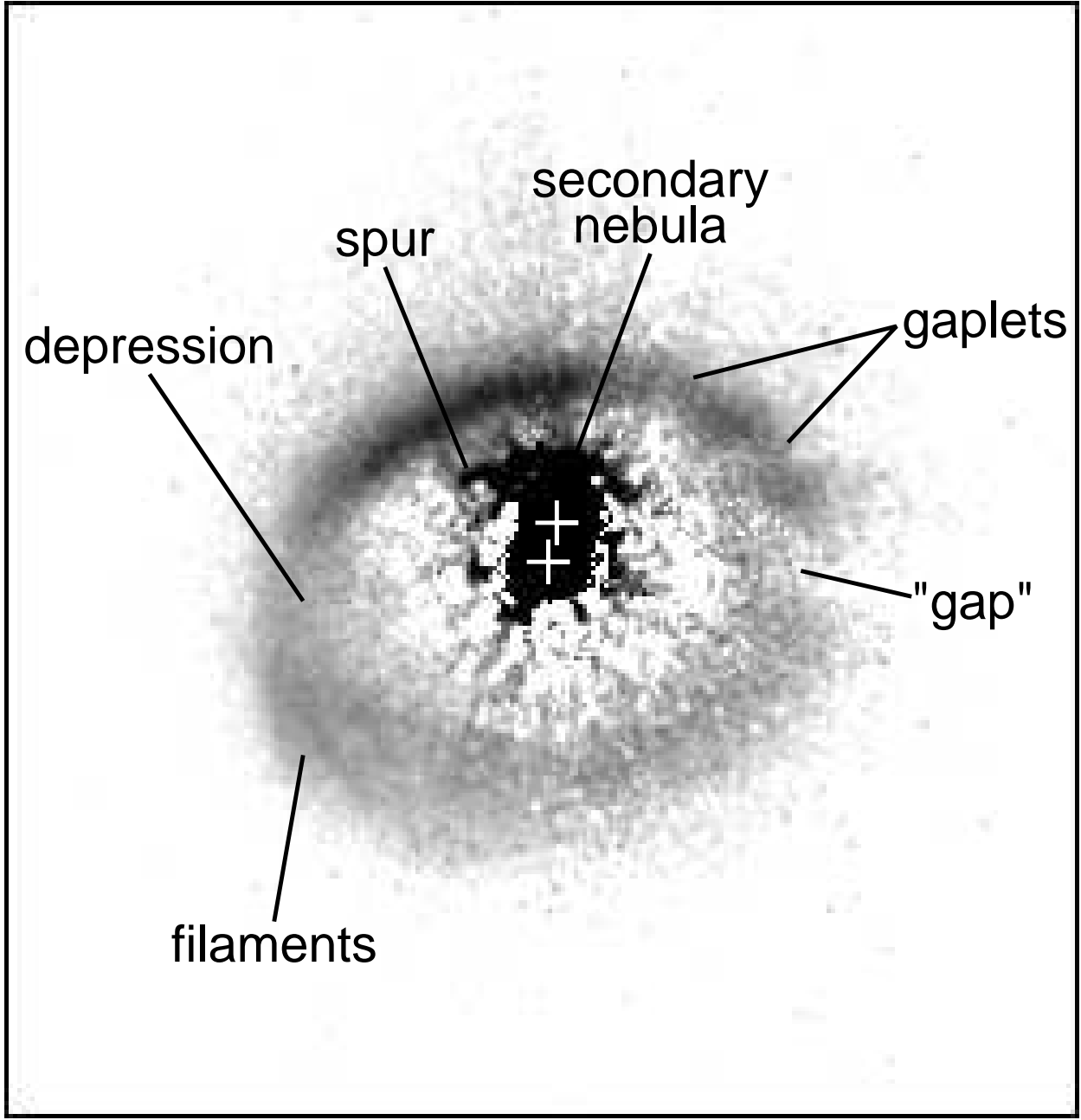}
\caption{ACS F814W PSF-subtracted image of GG Tau with features identified
(North is up).}
\label{fig3}
\end{figure}
 
There is a ``depression'' in the surface brightness along the eastern interior
edge (PA$=90^{\circ}-114^{\circ}$) relative to the adjacent sections of the
ring.  The depression seems more defined in $V$ than $I$, hinting of a possible
increase in extinction at shorter wavelengths by circumstellar material.
However, the red halo residuals prevent any definitive conclusions. 

The interior of the ring does not appear to be completely clear, and it is
brighter than the surrounding region immediately outside of the ring.  This is
especially apparent in the $V$ image, which is free of any red halo
contamination.  Excluding any small regions with significant subtraction
residuals near the stars that could bias the flux measurements, the interior
has median surface brightnesses of $V=20.1\pm0.4$ and $I_c=18.1\pm0.4$ mag
arcsec$^{-2}$.  Increasing the reference PSF intensity by 10\% during PSF
subtraction would eliminate this flux, but would lead to significant
oversubtraction outside of the ring.  This amount of adjustment is also greater
than the photometric measurement errors allow.  

As noted by KSW02 from the WFPC2 images, there appears to be a compact region
of reflecting material adjacent to the secondary star on its north side.  This
``secondary arc'' is clearly seen at the same position angle at both telescope
orientations in $I$, and at nearly the same position in both $V$ orientations.
The greater subtraction residuals at $V$ interfere with detail near the stars.
It therefore appears to be real and is not an instrumental artifact.  It
extends out to $\sim0\farcs 5$ from GG Tau Ab.  Its shape and brightness are
difficult to accurately determine given its proximity to the star.  At 0\farcs
3 from GG Tau Ab, its mean $I$ band surface brightness is $15.5\pm0.2$ mag
arcsec$^{-2}$.  This apparent nebulosity can also be seen in the NICMOS images
of Silber et al. (2000), though it was not identified or discussed there.  In
the NICMOS images of MDG02 it is located within the region of the diffraction
spikes and is not seen.  A short ``spur'' seems to extend from the nebulosity
and runs parallel to the forward edge.  This feature is seen at both
orientations at $I$, though it is more prominent at the second roll.  PSF
subtraction artifacts interfere with its detection at $V$.  There is no sign of
it in the WFPC2 images, but the NICMOS image of Silber et al. (2000) does
appear to show a similar feature in the same region.  This spur therefore seems
to be real and is probably not a PSF subtraction artifact.   The brightness of
the secondary nebula suggests that it is optically thick.  As discussed in
KSW02, it is too large to be a disk around the secondary and is too far from
the star to be material trapped at the system's Lagragian L2 point.  Its nature
is still unexplained.  Its brightness and size makes it a good target for
adaptive optics observations on future very large telescopes.  Perhaps with
additional resolution it will be possible to detect features such as spiral
arms that could point to a dynamical cause for its existence.

\subsection{Ring Photometry}

Azimuthal surface brightness plots for the combined ACS data for each filter
are shown in Figure 4, along with the WFPC2 $I$ band plot.  The tracer line
used for these was defined by the peaks of 3$^{rd}$-order polynomials that were
fit to the radial profiles extracted from the ACS $I$ band image.  The tracer
was manually adjusted to cross the ``gap'', where the fits failed to converge.
The brightness at each azimuth was measured using the median flux within a
0\farcs 075$^2$ box centered on line.  An ellipse fit through the tracer has a
semi-major axis of 1\farcs 47, minor axis position angle of 20$^{\circ}$, and
eccentricity of 0.63.  These values agree well with those derived by MDG02
using a similar method.

\begin{figure} 
\plotone{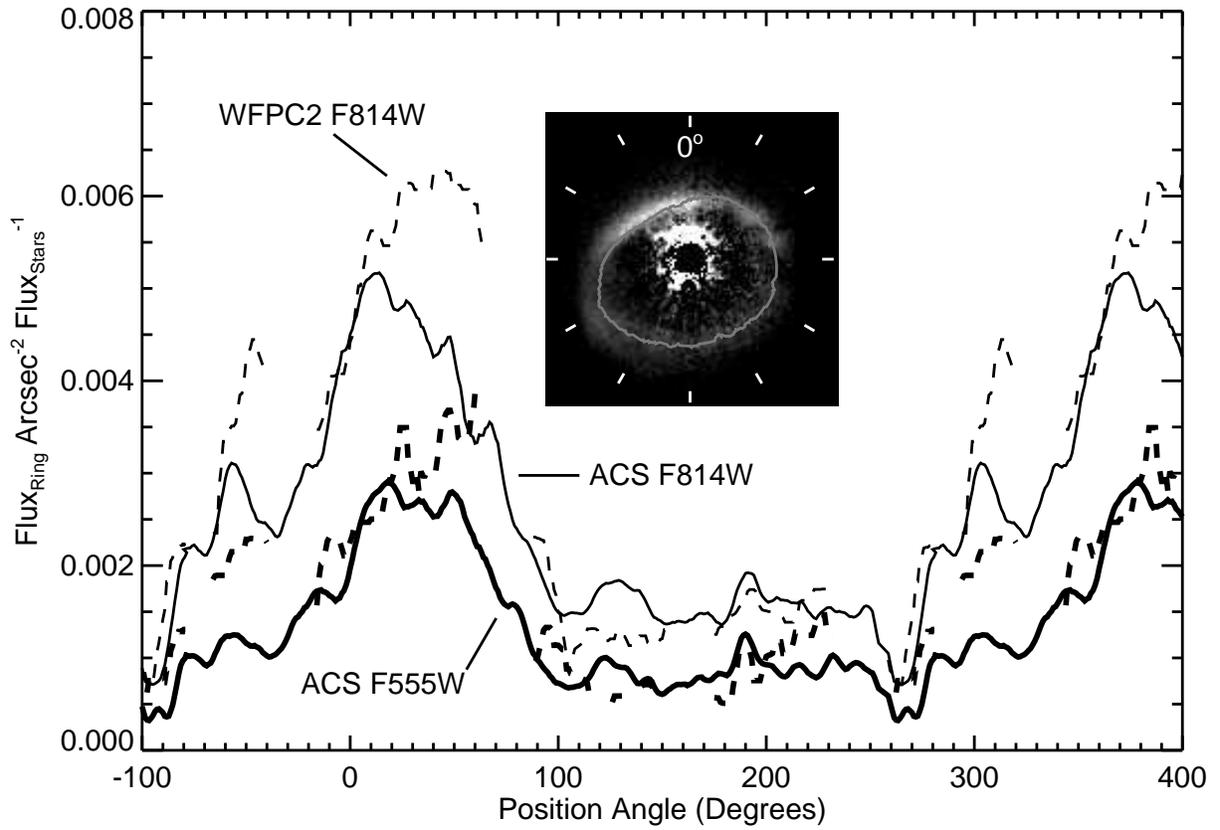} 
\caption{Azimuthal surface brightness profiles of the GG Tau ring as measured
in the ACS and WFPC2 $V$ and $I$ images. The data from the ACS roll- combined
images are shown. The bold dashed line is WFPC2 $V$. The values were measured
along the line traced along the ring as shown in the inset image.} 
\label{fig4} \end{figure}

Separate profiles (Figures 5 and 6) were extracted from the ACS images in each
filter and at each orientation to quantify the variations caused by PSF
subtraction errors, including the red halo asymmetry.  The $V$ band plots show
a $\sim15\%$ difference along the forward edge between the two orientations,
which appears to be caused by subtraction residuals, but the other sections
match to better than a few percent.  The effects of the red halo asymmetry are
apparent in the $I$ band plots.  They show that in the northwest quadrant
(PA=$250^{\circ}-350^{\circ}$) the ring surface brightness was $20\%-60\%$
greater at the first roll than the second.  The ring is otherwise $10\%-20\%$
brighter in the second-roll image in the north (PA=$10^{\circ}-70^{\circ}$) and
the south-southeast (PA=$110^{\circ}-180^{\circ}$).  The brightnesses are
similar in the southwest quadrant and in the eastern apex.  

\begin{figure}
\plotone{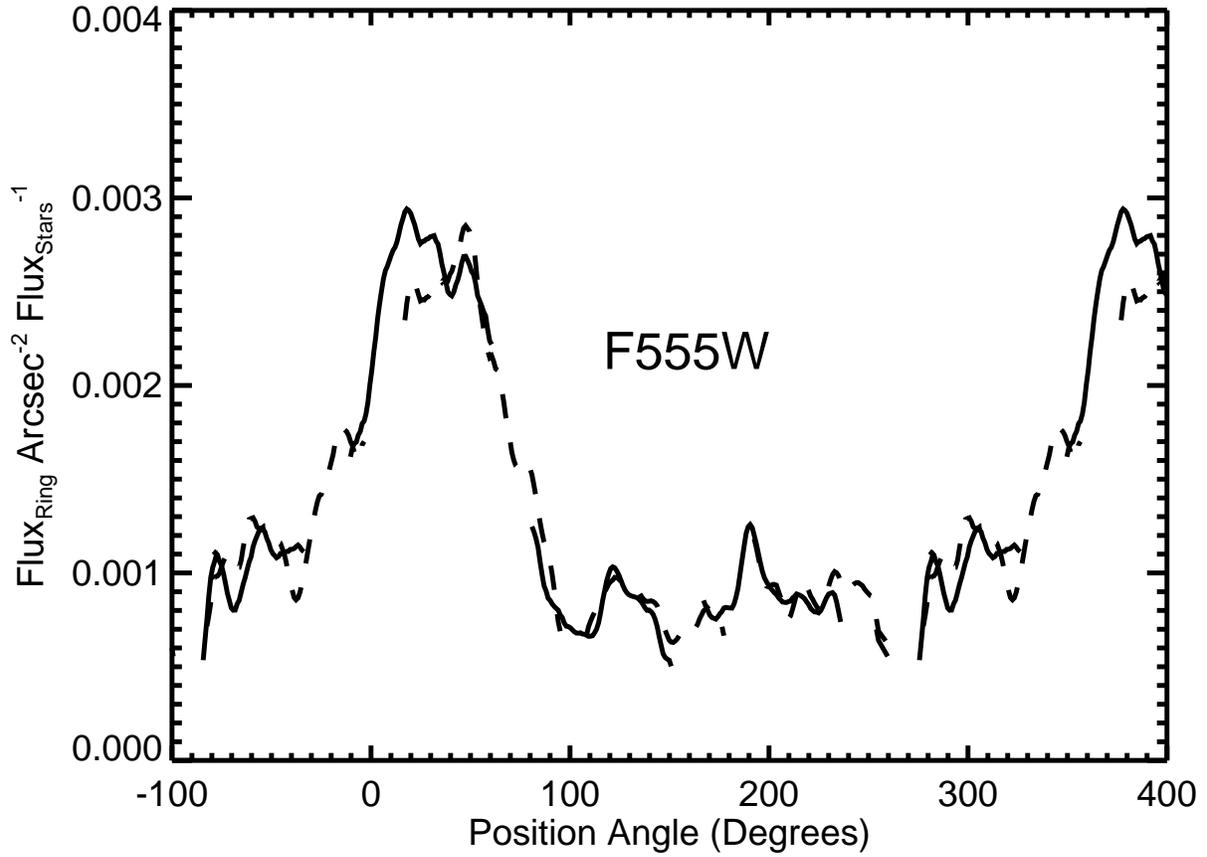}
\caption{Azimuthal surface brightness profiles of the GG Tau ring measured in 
the ACS $V$ image from each orientation (first orientation = solid, second = 
dashed).  Regions contaminated by diffraction spike residuals are not plotted.}
\label{fig5}
\end{figure} 

\begin{figure}
\plotone{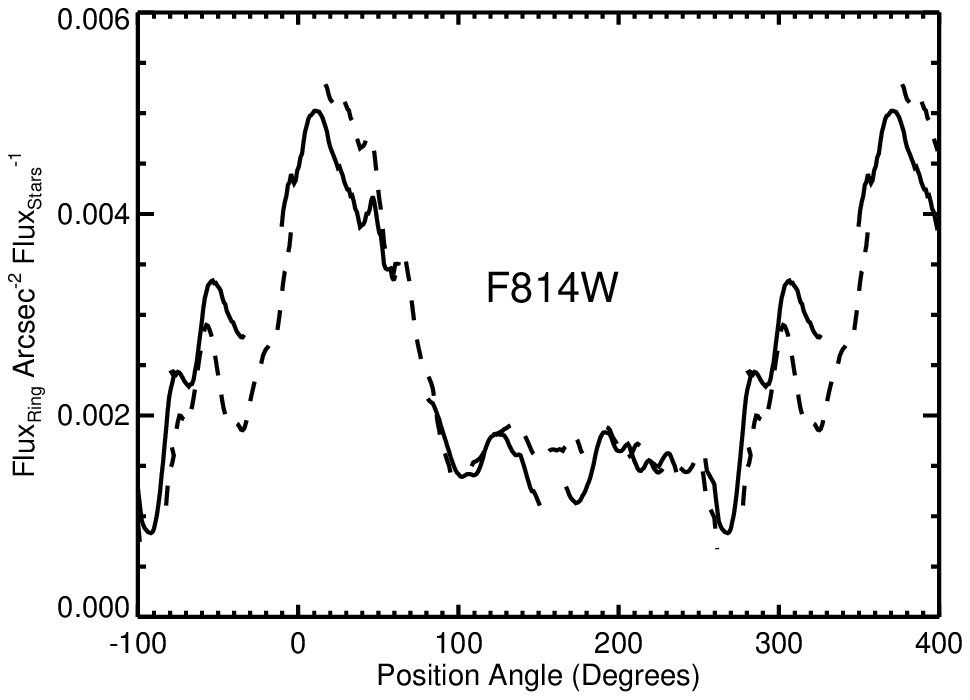}
\caption{Azimuthal surface brightness profiles of the GG Tau ring measured in 
the ACS $I$ image from each orientation (first orientation = solid, second = 
dashed).  Regions contaminated by diffraction spike residuals are not plotted.}
\label{fig6}
\end{figure} 

\begin{figure}
\plotone{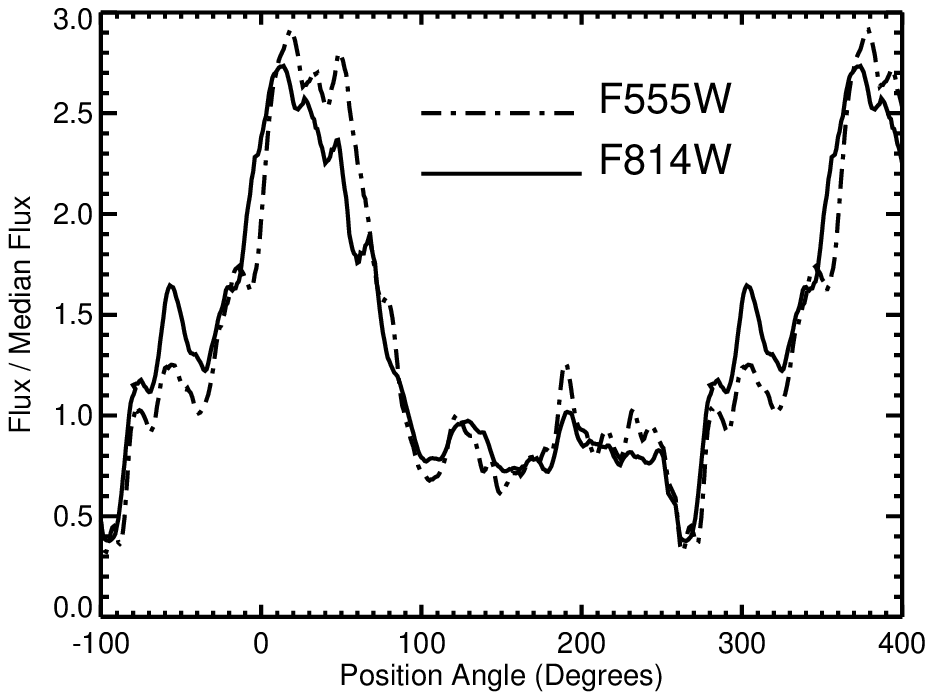}
\caption{Azimuthal surface brightness profiles of the GG Tau ring measured
in the roll-combined ACS $V$ and $I$ images, normalized by the median flux
within the ring in each bandpass.}
\label{fig7}
\end{figure} 

To allow quantitative comparisons with the D04 WFPC2 and
NICMOS results, fluxes in the ACS $V$, ACS $I$, and WFPC2 $I$ images were
measured with the same elliptical annular aperture that they used.  The
concentric ellipses defining the annulus had semi-major axes of 0\farcs 99 and
2\farcs 2, $e=0.65$, minor axis PA=20$^{\circ}$, and were concentric with the
ellipse fit through the tracer line.  Fluxes were measured separately in the
forward (PA=$275^{\circ}-95^{\circ}$) and back (PA=$95^{\circ}- 275^{\circ}$)
halves of the annulus.  

To exclude any contribution from the ``secondary arc'', that region was
manually masked out within the aperture.  The same area was essentially masked
by the diffraction spikes in the NICMOS images measured by MDG02 (and used by
D04 and so did not contribute to their values.  However,
their measurement of the WFPC2 image may be biased by it.    

To estimate the photometric errors introduced by the red halo asymmetry,
modified ACS $I$ band images were constructed for each roll.  Pixels containing
diffraction spikes at one roll were replaced by the corresponding,
uncontaminated pixels from the other roll.  This provided two separate images
of the entire ring.  Replacing the diffraction spikes with data from the other
roll should not greatly impact the photometry as the spikes cover a small total
area of the ring.   The same procedure was applied to the two $V$ images as
well.  The ring fluxes from these separate images were used to define the error
ranges in the flux ratios that follow.  In addition, for the purposes of the
flux ratio measurements only, the diffraction spike regions in the WFPC2 $I$
image were replaced with the corresponding values from the roll-combined ACS $I$
image. 

The fluxes of the ring (F$_{ring}$) relative to the combined stellar fluxes
(F$_{stars}$) are 0.58\% $\pm$0.01\% ($V$) and 1.09\%$\pm$0.03\% ($I$).  The
$I$ band value likely represents the upper limit as the residual red halo would
make the ring appear brighter than it actually is.  The integrated
forward/backside (F/B) flux ratios are 1.25 $\pm$0.01 ($V$) and 1.31$\pm$0.10
($I$).  This measurement is clearly sensitive to the halo asymmetry.  The WFPC2
$I$ band image values are F/B$=1.56\pm0.03$ and
F$_{ring}$/F$_{stars}$=1.23\%$\pm$0.02\%.  These can be compared to the
measurements of the same WFPC2 image by D04, who derived F/B=1.80$\pm$0.03 and
F$_{ring}$/F$_{stars}$=1.30\%$\pm$0.03\%.  These larger values may be due to
contamination by the secondary arc, which we avoided.

The ring is clearly redder in the visible than the measured light from the
stars, with a median color of $V-I_c=2.66$ compared to 1.85 for the combined
stellar fluxes.   This agrees with the WFPC2 color measurement of $V-I_c=2.7$
(KSW02).  We were unable to identify color variations along the ring that were
above the uncertainty levels set by the red halo residuals.  The data does
suggest that the ``depression'' in the east is slightly more pronounced in $V$
than $I$, but again that may be due to the residual red halo contributing to
the addition $I$ band flux.

\section{Discussion}

The red halo contamination of the ACS F814W images prevents reliable detections
of any low-level color variations within the ring.  Given the photometric
uncertainties, we limit our study to previously-seen and new structures, the
general color of the ring in the visible, and the apparent difference in the
forward edge brightness in the ACS and WFPC2 $I$ band images.

\subsection{The stars}

The 1994 and 1997 WFPC2 and 2002 ACS data provide a limited sample of resolved
visible-wavelength stellar photometry.  Among these epochs GG Tau Ab varied by
$\Delta V=0.61$ and $\Delta I=0.39$ while Aa changed by $\Delta V=0.11$ and
$\Delta I=0.07$.  Ab also had a larger variation in color ($\Delta V-I_c=0.31$
versus 0.06 for Aa).  Ab was most blue when it was brightest in $V$.  This may
be expected if the greater flux was due to increased accretion or an accretion
spot coming into view due to stellar rotation (prominent accretion emission
lines are seen in the $V$ filter).  This would suggest (ignoring the small
sample) that Ab is the more active star.  However, visible-wavelength {\it HST}
spectra of the system (Hartigan \& Kenyon 2003) indicate the opposite, unless
the spectra were taken while Ab was in a quiescent state or when an accretion
spot was not visible.  Variation caused by the orbital motion of circumstellar
material is a less likely explanation as Ab is not faintest in $V$ and $I$ at
the same time.

The ACS angle and separation are the same (within the errors) as those reported
by D04, whose well-resolved data were taken three months
later.  The PA errors for both measurements are larger than the 0.38$^{\circ}$
motion that would be expected between those epochs.  The {\it HST} and D04
results indicate that the binary separation has not changed between 1994 --
2002 within the measurement errors, with a mean of 0\farcs 251$\pm$0\farcs 003.
Tamazian et al. (2002) predicted a separation of 0\farcs 244 for the epoch of
the ACS observations, outside of the ACS measurement errors.  Their predicted
position angle is consistent with that observed, however.  Their orbit may be
biased by speckle results, which appear to show large and inconsistent
variations in both PA and separation among the various epochs.

\subsection{Comparisons with Similar Wavelengths}

Comparisons of the ACS and WFPC2 $V$ band images do not reveal significant
differences given the large PSF subtraction residuals and low exposure depth of
the WPFC2 data.  The general appearance is similar, but details such as the
``depression'' are poorly defined.  

The ACS and WFPC2 $I$ band images are more comparable, though care must be
taken to identify possible red halo effects.  The general appearance is the
same between them.  The ``gap'' and secondary arc are both present, though the
``gaplets'' are not clearly present in the WFPC2 image (saturated column
bleeding and diffraction spike residuals are present in the WFPC2 data).  The
``depression'' is seen in the WFPC2 image, but it is not quite as well defined
as it is in the ACS one.  

The forward side does appear to be $\sim20\%$ brighter in WFPC2 than in ACS
while the back sides are fairly equal.  The integrated F/B ratios are 1.56
(WFPC2) and 1.31 (ACS), and the F/B amplitude ratios are $\sim4$ (WFPC2) and
$\sim3.3$ (ACS).  These discrepancies are above the uncertainty level expected
from the red halo contamination in the ACS images. The back side of the ring
has essentially the same brightness in both ACS and WFPC2 images.  We believe
that this is a real effect and indicates that some variability in the
illumination pattern is likely.  We note that there does not appear to be any
significant changes in the positions of any features between the epochs,
including the ``gap'', indicating that any shadowing material causing it must
be further from the stars than whatever is altering the pattern on the forward
side.   

Time-dependent variations in the appearances of reflection nebulae (envelopes,
disks) surrounding young stars have been seen in R Mon (Lightfoot 1989), HH 30
(Stapelfeldt et al. 1999), and others.  The timescale of these changes ({\it
e.g.} 5 years for GG Tau) is too short to be caused by physical variations in
the ring itself -- they must be due to irregular illumination by the stars.
Continuum emission from accretion spots on the star could increase the amount
of light seen by a particular section of the disk, and as the star rotates they
could vary the illumination pattern.  Another possibility is shadowing from
irregularly distributed circumstellar material near enough to the star to have
a short orbital period or be placed where the orbital motion of the stars
alters the projected light.  The variation in the brightness of the near edge
between the two epochs matches the change in the observed flux from GG Tau Ab.
This suggests that if light from Aa was blocked by a circumstellar disk in the
ring plane, Ab could be the primary illumination source for this edge.  A final
potential cause is a light echo from an intense, short-term brightening of one
of the stars.  Due to the light travel time, the starlight reflected by the
back side of the ring is seen about 1.3 days later than that from the forward
edge.  Differences in the phases of the stellar light curves might then result
in equal backside ring brightnesses at both epochs while the forward side
appears brighter at one of them.  This would require either star to vary by
more than twice its observed intensity range over a one day timescale.

\subsection{Comparisons with Different Wavelengths}

As noted before, direct comparisons of the $V$ and $I$ ACS images are difficult
because of the PSF subtraction residuals in $V$ and the red halo in $I$.  The
profiles in Figure 7 would suggest that the forward edge in $V$ is somewhat
brighter than in $I$, relative to the rest of the ring.  However, this
difference is within the range of errors expected from the residuals.  

The NICMOS image of Silber et al. (2000) appears very similar to those from
ACS.  The ``gap'' is clearly seen but not the ``gaplets.'' The ``depression''
in the east is also apparent, but the outer edge there is brighter compared to
the ACS images.  Silber et al. report a $\sim4.0\times$ difference between the
front and back side surface brightnesses, compared with $\sim3.3\times$ seen in
the ACS image.  They note the presence of two compact features within the clear
region that they suggest may be reflecting material.  The first is an elongated
streak to the southeast of the star.  A similar feature is not seen in the
other images, except perhaps in the MDG02 $J$ one, in which a thin residual
streak extends along the same direction.  However, this streak appears sharper
than the PSF resolution of the system and is probably a chance subtraction
artifact.  The Silber et al.  artifact may also be a subtraction residual or
polarizer ghost.  The second feature looks very much like the ``spur'' seen
extending from the secondary arc in the ACS images.

The MDG02 NICMOS F110W ($\sim J$), F160W ($\sim H$), and F205W ($\sim K$)
images have considerably poorer sampling than those from ACS (0\farcs 076 vs.
0\farcs 025 pixel$^{-1}$).  The F205W image was taken at a non-optimal focus
position and so is ignored.  The residuals from the diffraction spikes, which
appear $2\times$ wider at $H$ than $I$, block large portions of the ring
including the ``gap'', ``depression'', and secondary arc.  The two ``gaplets''
are visible, however.   The front/back amplitude ratios are $\sim4$ for both
filters, compared to $3.3$ for ACS $I$.  To measure the integrated ring fluxes,
they replaced the blocked pixels of the ring with the mean disk flux per pixel.
Their integrated F/B ratios are $1.41\pm0.03$ ($J$) and $1.39\pm0.02$ ($H$),
compared with 1.31 for ACS $I$.  These images were taken only 14 days after
those from WFPC2, but the WFPC2 F/B ratio is considerably higher ($1.56-1.80$,
depending on the study).  How much of this difference can be attributed to the
longer wavelength is unknown, but this may be evidence for short-period ring
illumination variations.

The Keck 3.8 $\mu$m ($L'$) image (D) has some significant differences in
comparison with those at shorter wavelengths.  The distance between the
binary's center of mass and the forward edge of the ring is 0\farcs 1 greater
than in the {\it HST} images.  D04 attribute this to the decreased opacity at
longer wavelengths of the dust along the inner forward edge.  The ``gap'' is
seen, though at reduced contrast relative to the other {\it HST} and
ground-based images, perhaps indicating a reduction in the extinction caused by
the shadowing circumstellar material.  There are no signs of the ``gaplets.''
The eastern side is well defined in $L'$, while it is ambiguous in ACS $I$.
This may also point to reduced extinction.  The southeast quadrant is the
faintest section in the Keck image, being within the noise; in ACS, this
section is brighter than the eastern extreme, which contains the
``depression.''  There is no indication of the ``secondary arc'' above the
level of the residuals near the stars. 

The most significant feature of the Keck $L'$ image is the unexpectedly large
amount of forward scattering.  The F/B amplitude ratio is $>5$ and the
integrated F/B ratio is $\sim2.5$, well above that seen in any of the shorter
wavelength images.  A single ISM-like grain size distribution would cause
scattering to become more isotropic at longer wavelengths, in which case the
$L'$ image should show little forward scattering, extrapolating from the
shorter wavelengths.  D04 suggest that Keck is seeing larger ($>1\mu$m)
particles that are not visible at shorter wavelengths, thus explaining the
increased forward scattering.  These larger particles are located closer to the
disk midplane ($\sim$25 AU versus $\sim$50 AU at shorter wavelengths) and so
could only be seen at longer wavelengths that could penetrate through the thick
dust.  That the $L'$ image is seeing further into the disk is evident by the
increase in the separation of the forward ring edge and the stars.  D04
suggest that this vertical stratification of particle sizes relative to the
disk height is evidence for grain growth close to the disk midplane, where
particles have a greater chance of colliding and coagulating.

\subsection{Shadows and Warping}

The variations along the ring suggest that the illuminated surface of the disk
is irregular.  They include:

\begin{itemize}

\item
The intensity distribution along the forward edge is asymmetric about the
line of sight.

\item
The forward side appears to vary in brightness over time.

\item 
The western side contains a series of likely azimuthal shadows (the ``gap'' and
``gaplets'').

\item
The western side of the ring is brighter, narrower, and has a sharper (though 
irregular) inner edge than the eastern side.

\item 
The extreme eastern inner edge has a lower intensity and the outer edge contains 
narrow filaments.

\end{itemize}

The first three of these can be easily explained by shadowing caused by
optically thick (or nearly so) concentrations of dust between the stars and the
disk.  The asymmetry in the forward edge could be altered by material very near
one of the stars, perhaps associated with the secondary arc.  Orbital motion of
the material or the star could then change the distribution of light along the
forward surface over a matter of hours or days.  

The gaps along the western side are almost certainly shadows.  There is no sign
of a depletion of material in the gap in the millimeter images (Guilloteau,
Dutrey, \& Simon 1999).  Given the viscosity of the massive, gas-rich disk, it
is likely that any resonantly-created gap would be filled in quickly.  The
shadowing material must be fairly dense, though, to cause the most prominent
gap to appear even at $\lambda=3.8\mu$m.

The last two items in the list are more difficult to explain with only
shadowing, and they imply that the ring is intrinsically asymmetrical.
Guilloteau, Dutrey, \& Simon (1999) noted that their mm continuum image shows
an apparent density enhancement along the western side, which the scattered
light images show is brighter than the opposite side.  If this enhancement is
real, then it may indicate that the density distribution of scatterers is also
different around the ring.  An azimuthal dependence of the disk flaring or
scale height would alter both the scattering phase angle and the projected
surface area.  A change in the flaring function could explain why the east side
appears wide and poorly-defined along its inner edge while the west side is
bright, narrow, and has a well-defined edge.  It could also explain the
asymmetry along the forward edge.  However, the timescale for density
distribution variations is too long to cause the illumination changes seen
between the ACS and WFPC2 images.  

Variations in the density distribution would not be surprising, given the
strong tidal influence of the binary.  Most circumbinary disk models
(Artymowicz \& Lubow 1996) show density asymmetries caused by the orbital
eccentricity of the binary, including spiral patterns.  These could cause the
scattering surface of the disk to be highly variable on small spatial scales,
in effect creating localized flared or self-shadowing regions.  Perhaps this is
an explanation for the filaments.

\subsection{Ring Color}

The ring has a significant red excess in the visible ($V-I_c=0.8$) while the
near-IR colors are more neutral ($J-H=0.1$ and $H-K=-0.04$; MDG02).  The
easiest explanation for this is that increasing extinction from circumstellar
material accounts for the lower F$_{ring}$/F$_{stars}$ flux ratio towards
shorter wavelengths.  KSW02 derived a required extinction of $A_v$=1.2 mag.
This simple solution, however, ignores any color effects caused by the
wavelength-dependent scattering properties of the ring material itself.  

Wood et al. (1999) reported the first results of matching multiple scattering
models of GG Tau to multi-color data (the $J$, $H$, and $K$ band images of
R96), assuming grain properties that reproduce the ISM extinction curve.  Their
models resulted in a blue, rather than red, color excess, even when extinction
by a  circumstellar disk was added to match F$_{ring}$/F$_{stars}$.  MDG02
later computed models to match their NICMOS images.  Using a somewhat similar
disk geometry but different grain properties (though still matching the ISM
extinction curve), they were able to produce a significant red excess without
extinction by any circumstellar material.  However, their models differ from
their NICMOS data by significant amounts in terms of F$_{ring}$/F$_{stars}$ and
F/B ratios.  Their results suggest that uncertainties in grain properties (i.e.
scattering phase function, albedo, and size distribution) may be the primary
source of color mismatches between models and observations, rather than
geometrical or possible circumstellar extinction effects.

\subsection{Ring Variations and Modeling}

D04 have expanded on the work of MDG02 to include models for the WFPC2 $I$ band
and Keck $L'$ images in addition to those from NICMOS.  Rather than simply
matching the integrated F/B ratios, they attempted to find models whose
azimuthal brightness profiles best matched the observed ones.  They claim that
this provides a more accurate indication of the grain scattering properties
than does the simple F/B contrast ratio.  As previously discussed, they were
unable to match the azimuthal profiles at all wavelengths with a single grain
size distribution, requiring instead a separate distribution of larger grains
in order to explain the large amount of forward scattering at $L'$.  Their
models, however, are incomplete as they significantly underestimate the F/B
contrast and overestimate the F$_{ring}$/F$_{stars}$ ratio at every wavelength
(they do suggest $A_v$=1 mag of extinction between the stars and ring to reduce
the ratio).  So far, no one has developed a model that simultaneously
reproduces the ring's color, forward/backside contrast, and ring-to-stars flux
ratio.

The apparent position-and-time-dependent variability in the illumination
pattern on the ring may create difficulties for those modeling the dust
distribution and grain properties of the GG Tau disk using non-contemporaneous
datasets, especially when comparing images at multiple wavelengths.  The
forward/backside contrast is the strongest constraint on the scattering phase
function, and it is now evident that this value can vary over time within the
same passband.  Time-resolved imaging of the GG Tau ring is required to verify
the illumination variations and establish the variation timescale.  This may be
possible in the future with further improvements to ground-based adaptive
optics or in the present with additional {\it HST} observations.    

The need obviously exists for integrated modeling that produces a dynamical
simulation of the disk and then shows how it would appear in scattered light
for a series of assumed grain and initial dust and gas distributions.  Such
models could indicate how much asymmetry would be expected from disk surface
warping.

\section{Conclusions}

We have have presented {\it HST}/ACS $V$ and $I$ band images of the GG Tau
binary system and its disk taken in 2002.  They confirm the existence of
material near the secondary component, filamentary or spiral structure along
the eastern side of the ring along with a broad region of low surface
brightness (the ``depression''), and the ``gap'' in the west as well as less
prominent gaps toward the northwest.  The $V$ image unambiguously shows for the
first time the far side of the circumbinary ring.  The ring is redder than the
apparent light from the stars with $\Delta V-I_c=0.81$.  There are no
significant color variations within the disk.  The forward edge of the ring in
the 1997 WFPC2 $I$ image was $\sim20\%$ brighter than it was in the ACS
exposure, indicating a time-variable illumination pattern.

The gaps indicate that there is circumstellar material that creates shadows on
the circumbinary disk.  The asymmetries in the ring suggest that the
illuminated disk surface may also be warped due to tidal interactions with the
binary.  There is a need for scattered light models using dust density
distributions derived from dynamical simulation of systems similar to GG Tau.
The models need not exactly match the observed images, but rather indicate how
warping of the disk may affect its appearance, placing some constraints on what
features may be due to illumination or structural variations.  The apparent
time variation in the illumination pattern further complicates the analyses,
especially when comparing images at multiple wavelengths taken at different
epochs.  Monitoring of GG Tau on short and long timescales is needed.
 
\section{Acknowledgements} 

ACS was developed under NASA contract NAS 5-32865.  This research was supported
by the ACS Science Team under NASA grant NAG5-7697.  The authors thank F.
M{\'e}nard and C. McCabe for discussions concerning these and previous results,
and the reviewer for many useful suggestions.

\end{document}